\documentclass[prl,showpacs,amsmath,amssymb]{revtex4}
\usepackage{graphicx}% Include figure files
\usepackage{dcolumn}% Align table columns on decimal point
\usepackage{bm}% bold math
\usepackage{natbib}

\newcommand{\al}{\ensuremath{\alpha}}
\newcommand{\dal}{\ensuremath{\Delta \alpha/ \alpha}}
\newcommand{\gp}{\ensuremath{{g}_{p}}}

\newcommand{\y}{\ensuremath{{m}_{e}/{m}_{p}}}

\newcommand{\beq}{\begin{equation}}
\newcommand{\eeq}{\end{equation}}
\newcommand{\noi}{\noindent}
\newcommand{\lb}{\left(}
\newcommand{\rb}{\right)}
\newcommand{\lsb}{\left[}
\newcommand{\rsb}{\right]}

\begin{document}

%\preprint{APS/123-QED}

\title{Constraining the variation of fundamental constants using 18cm OH lines}
\author{Jayaram N. Chengalur}
\email{chengalur@ncra.tifr.res.in}
\affiliation{ NCRA/TIFR, P. O. Bag 3, Ganeshkhind, Pune 411007, India}
%\altaffiliation[Also at ]{Physics Department, XYZ University.}
\author{Nissim Kanekar}
\email{nissim@astro.rug.nl}
\affiliation{ Kapteyn Institute, Groningen University, The Netherlands}

\date{\today}% It is always \today, today,

\begin{abstract}
We describe a new technique to estimate variations in the fundamental 
constants using 18cm OH absorption lines. This has the advantage that all lines 
arise in the same species, allowing a clean comparison between the measured 
redshifts. In conjunction with one additional transition (for example, an 
HCO$^+$ line), it is possible to simultaneously measure changes in $\alpha$, 
$g_p$ and $y \equiv m_e/m_p$. At present, only the 1665~MHz and 1667~MHz lines 
have been detected at cosmological distances; we use these line redshifts in 
conjunction with those of HI 21cm and mm-wave molecular absorption in a 
gravitational lens at $z\sim 0.68$ to constrain changes in the above three 
parameters over the redshift range $0 < z \lesssim 0.68$. While the 
constraints are relatively weak ($\lesssim$ 1 part in $10^3$), this is the 
first simultaneous constraint on the variation of all three parameters. 
We also demonstrate that either one (or more) of $\alpha$, $g_p$ and 
$y$ must vary with cosmological time or there must be systematic velocity 
offsets  between the OH, HCO$^+$ and HI absorbing clouds.
\end{abstract}

\maketitle

\section{\label{sec:intro} Introduction}

The recent claim by Webb et al. \cite{webb99,webb01} that the fine structure constant 
$\alpha$ evolves with redshift, with $\dal = (-1.88 \pm 0.53) \times 10^{-5}$ from 
$z \sim 1.6$ to today and $\dal = (-0.72 \pm 0.18) \times 10^{-5}$ for $0.5 < z < 3.5$~
(but see \cite{bekenstein03}) has spurred interest in the possibility that the numerical 
values of the fundamental constants change with time. Theories that can account for 
such variations include extra-dimensional Kaluza-Klein theories and superstring 
theories. In such models, the values of the coupling constants depend on the 
expectation values of some cosmological scalar field(s); changes in the values 
of the coupling constants are thus to be expected if this field varies 
with location and time. Further, depending on the details of the theory, all of 
the different coupling constants (such as $\alpha$, the proton g-factor $g_p$, 
the electron-proton mass ratio $y \equiv m_e/m_p$, the gravitational constant $G$, etc)
could, in principle, vary simultaneously.
For example, Calmet~\&~Fritzsch \cite{calmet02} and Langacker~et al. 
\cite{langacker02} find that variations in the value of $\alpha$ should be 
accompanied by much larger changes (by $\sim 2$ orders of magnitude) in the 
value of $y$. However, Ivanchik~et al.  \cite{ivanchik03} constrain the variation 
in $y$ to be $(3.0 \pm 2.4) \times 10^{-5}$ over the redshift range  
$ 0< z <3$, comparable to the change claimed in the fine structure constant.
A review of the available experimental and observational measurements 
on the variability of the coupling constants can be found in \cite{uzan03}.

One of the main problems in most of the astrophysical techniques used to measure (or 
constrain) the values of the different constants (e.g. \cite{carilli00,webb01,ivanchik03})
is that they involve a comparison between the redshifts of spectral lines of
different species (e.g. the HI 21cm line, millimetre-wave molecular lines 
and optical fine structure lines \cite{webb01,carilli00}). These species are 
unlikely to all arise at the same physical location in a gas cloud and might thus 
have systematic velocity offsets relative to each other; the redshift differences 
may thus be dominated by these effects rather than the measurement errors
(i.e. the spectral resolution, which can be quite small, $\sim$~few km/s, for 
HI 21cm and mm-wave molecular absorption spectra). Conclusions drawn from a 
comparison between different species might thus well be incorrect.

Clearly, the best way to test the variation of the coupling constants is to use 
lines originating from a {\it single species}, but with different 
dependences on these constants. Further, since many constants may be varying 
simultaneously, it would be very useful if one could simultaneously measure the 
changes in a number of constants from this single species, rather than assuming 
that changes occur in only one of the constants and that the others remain unchanged.
We present here a new technique which satisfies both these requirements, using 
the 18cm lines of the OH radical.

	The ground $^2\Pi_{3/2}$ J=3/2 state of OH is split into two 
levels by $\Lambda$-doubling and each of these $\Lambda$-doubled levels 
further split into two hyperfine states. Transitions between these levels 
lead to four spectral lines with wavelength $\sim 18$cm. Transitions with 
$\Delta F=0$ are called the ``main'' lines, arising at rest frequencies 
of 1665.4018~MHz and 1667.3590~MHz, while transitions with $\Delta F = 1$ 
are called ``satellite'' lines, with rest frequencies of 1612.2310~MHz and 
1720.5299~MHz. Since the four OH lines arise from two very different physical 
processes, viz. $\Lambda$-doubling and hyperfine splitting, the transition 
frequencies have different dependences on the 
fundamental constants. A perturbative treatment of the OH molecule has been 
carried out by Dousmanis et al. \cite{dousmanis55} (see also \cite{townes55});
we use the expressions derived in these references to determine the dependence
of linear combinations of the four OH line frequencies on $\alpha$, $g_p$ and 
$y \equiv m_e/m_p$ and show that it is possible to simultaneously measure 
changes in both $\alpha$ and $y$, if all four line frequencies are known 
(assuming that $g_p$ does not vary with time). Since OH and HCO$^+$ column densities
are observed to be tightly correlated, both in the Galaxy \cite{liszt96} and out to $z
\sim 1$ \cite{kanekar02}, these species are likely to arise at the same physical 
location; a comparison between the redshifts of the four 18cm OH lines and 
the HCO$^+$ line should thus allow one to constrain the evolution of all the 
above three parameters. We use our observations of the OH main lines 
in the $z \sim 0.6846$ gravitational lens towards B0218+357, in tandem with 
published HI 21cm and mm-wave molecular redshifts, to constrain $\Delta y/y$ 
between $z \sim 0.68$ and today. Finally, as this work was being written up, 
an analysis on the use of OH lines to constrain changes in fundamental constants 
was also carried out by Darling \cite{darling03}; the latter, however, only 
considers variations in the fine structure constant $\alpha$.

\section{ Constraints from Radio Spectral Lines}

Consider two transitions whose rest frame frequencies $\nu_i(z), i = 1, 2$ 
depend on redshift, due to the evolution of various fundamental constants, 
such as $\al, \gp, \y$, etc. If the lines arise in a source at a ``true'' 
redshift $z$, the {\it measured} redshift $\hat z_i$ of each line is given by

\beq
\lb 1 + {\hat z_i} \rb^{-1} \;\;=\;\; \frac{\nu_i(z) }{ \lb 1 + z \rb \nu_i(0) }
\;\;=\;\; \frac{1 + \lsb \Delta \nu_i(z)/\nu_i(0) \rsb }{1 +z} \;\; ,
\eeq
\noi where $\Delta \nu_i (z) \equiv \nu_i(z) - \nu_i(0)$; note that $\Delta 
\nu_i (z) = 0$ if the fundamental constants do not change with time. The 
first order difference between the two measured redshifts $\Delta z = 
\hat{z}_1 -\hat{z}_2$ is then 

\beq 
\label{eqn:redshift}
\frac {\Delta z }{ 1 + {\bar z} }\;\; =\;\; \lsb \frac{ \Delta \nu_2}{\nu_2(0)} \rsb - 
\lsb \frac{\Delta \nu_1}{\nu_1(0)} \rsb \;\; ,
\eeq

\noi where $\bar z$ is the mean measured redshift. Given two spectral lines (or 
linear combinations of line frequencies) with different dependences on some 
fundamental parameter, one can thus use the differences between the measured 
redshifts to constrain the evolution of the parameter in question. 

In the case of the four 18cm OH lines, the following three independent relations 
have been shown to be satisfied by the line frequencies \cite{townes55,dousmanis55}; 
note that the lines must also satisfy the constraint $\nu_{1665} + \nu_{1667} 
= \nu_{1720} + \nu_{1612}$ 

\beq
\label{eqn:sum1}
\nu_A\;\; \equiv\;\; \nu_{1667} + \nu_{1665}\;\; =\;\; q_\Lambda\lsb 
\lb 2 + \frac{A'}{B'} \rb \lb 1 - \frac{2 - A/B}{X} \rb -
\frac{12}{X} \rsb \;\; ,
\eeq
\beq
\label{eqn:diff1}
\nu_B\;\; \equiv\;\; \nu_{1667} - \nu_{1665}\;\; =\;\; \frac{8d \lb X - 2 + A/B \rb }{15 X}  \: \: \:\:\:\: \mathrm{and}
\eeq
\beq
\label{eqn:diff2}
\nu_C\;\; \equiv\;\; \nu_{1720} - \nu_{1612}\;\; =\;\; 
\frac{4}{15X} \lsb 2a \lb 2X + 2 - A/B \rb  +  12b +\lb b+c \rb 
\lb X + 4 - 2A/B \rb \rsb
\eeq

\noi Equations (\ref{eqn:sum1}), (\ref{eqn:diff1}) \& (\ref{eqn:diff2}) correspond to the 
energy split due to $\Lambda$~doubling and the difference and sum of the hyperfine splits 
in the two $\Lambda$~doubled levels respectively. Here, $X \equiv \lsb \lb 
A/B \rb \{ {\lb A/B\rb} - 4 \}+ 16 \rsb^{1/2}$, 
$A$ is the fine structure interaction constant, $B$, the rotational constant, 
$A'$ and $B'$, the off-diagonal matrix elements of these operators, 
$q_\Lambda \approx 4B^2/h\nu_e$ (where $h\nu_e$ is the energy difference between 
the ground and first excited electronic state), and, finally, $a$, $b$, $c$ and $d$ 
are ``hyperfine constants'' \cite{dousmanis55}, whose experimental values are 
$a = 86.012 \pm 0.002$~MHz, $ b = -116.719 \pm 0.008$~MHz, $c   = 130.75 \pm 0.01$~MHz 
and $d = 56.632 \pm 0.004$~MHz \cite{coxon79}. Numerically, $A/B = -7.547$ and 
$A'/B' = -6.073$ \cite{townes55}. These quantities have the following dependences on 
the fundamental constants $\alpha$, $y$, $R_\infty \equiv m_e e^4/\hbar ^3 c$ and $g_p$ : 
$A' \propto A \propto \alpha^2 R_\infty \;\;$, $B' \propto B \propto y R_\infty \;\;$ 
and $a, b, c, d \propto g_p \alpha^2 y R_\infty$ \cite{townes55}. For the rotational constant $B$, 
we have assumed, as usual (e.g. \cite{murphy01}), that variations in $(m_p/M)$, 
which are suppressed by a factor $m_p/U \sim 100$ (where $M$ is the reduced mass and 
U the binding energy) can be ignored. Thus, we have 
$ \lsb A'/B' \rsb \propto \lsb A/B \rsb \propto \lb \alpha^2/y \rb $.
Replacing the above scalings in equation~(\ref{eqn:sum1}) for $\nu_A$, we obtain
$\nu_A \propto y^2 R_\infty F\lb \alpha^2/y\rb \;\;$, where $F \equiv F \lb \beta \rb $ 
is a function which depends only on the ratio $\beta \equiv A/B \propto \alpha^2/y$ 
and is defined by 
\beq
F\lb \beta \rb = \lsb \lb 2 + \frac{6.073}{7.547}\beta \rb \lb 1 + 
\frac{2 - \beta}{X\lb\beta\rb} \rb + \frac{12}{X\lb \beta\rb} \rsb
\eeq
\noi Thus,
\begin{eqnarray}
\frac{\Delta \nu_A }{\nu_A} &=& 2\frac{\Delta y}{y} + \frac{\Delta R_\infty}{R_\infty}
+ \frac{\Delta F \lb \beta \rb} {F\lb\beta \rb}\\
&=& 2\frac{\Delta y}{y} + \frac{\Delta R_\infty}{R_\infty} 
+ \frac{\beta}{F} \frac{dF}{d\beta} \lsb 2 \frac{\Delta \alpha}{\alpha} - 
\frac{\Delta y}{y} \rsb 
\end{eqnarray}

\noi Evaluating the quantity on the right-hand-side of the above equation, we 
obtain 
\beq 
\label{eqn:temp1}
\frac{\Delta \nu_A }{\nu_A} = 2.571 \frac{\Delta y}{y} - 
1.141 \frac{\Delta \alpha}{\alpha} + \frac{\Delta R_\infty}{R_\infty}
\eeq

\noi In similar fashion, equations~(\ref{eqn:diff1}) and (\ref{eqn:diff2}) yield 
\beq 
\label{eqn:temp2}
\frac{\Delta \nu_B }{\nu_B} = 2.442\frac{\Delta y}{y} -
0.883 \frac{\Delta \alpha}{\alpha} + \frac{\Delta R_\infty}{R_\infty}
+ \frac{\Delta g_p}{g_p} \:\:\:\:\:\: \mathrm{and}
\eeq
\beq
\label{eqn:temp3}
\frac{\Delta \nu_C }{\nu_C} = 0.722\frac{\Delta y}{y} +
2.557\frac{\Delta \alpha}{\alpha} + \frac{\Delta R_\infty}{R_\infty}
+ \frac{\Delta g_p}{g_p}
\eeq

\noi Equations~(\ref{eqn:temp1} -- \ref{eqn:temp3}) govern the way in which 
a change in one of the fundamental constants affects the line rest 
frequencies. Combining them in pairs in equation~(\ref{eqn:redshift}) yields :

\beq
\label{eqn:z12}
\frac {\Delta z_{AB} }{ 1 + {\bar z_{AB}} } = 
\lsb \frac{\Delta \nu_B}{\nu_B} - \frac{\Delta \nu_A}{\nu_A} \rsb 
= -0.129 \frac{\Delta y}{y} + 0.258\frac{\Delta \alpha}{\alpha} 
+ \frac{\Delta g_p}{g_p} \;\; ,
\eeq

\beq
\label{eqn:z13}
\frac {\Delta z_{AC} }{ 1 + {\bar z_{AC}} } = 
\lsb \frac{\Delta \nu_C}{\nu_C} - \frac{\Delta \nu_A}{\nu_A} \rsb 
= -1.849\frac{\Delta y}{y} 
+ 3.698\frac{\Delta \alpha}{\alpha} + \frac{\Delta g_p}{g_p} \:\:\:\:\:\: \mathrm{and}
\eeq

\noi If all four OH lines are detected in absorption in a single cosmological 
system, we thus have two independent equations relating the differences in measured 
redshifts to the changes in $\alpha$, $g_p$ and $y \equiv (m_e/m_p)$. Most 
analyses (using other lines) assume that $g_p$ remains unchanged and then estimate
the variation of $\alpha$ \cite{murphy01,carilli00}. If one makes the same assumption
in the case of the OH lines,
equations~(\ref{eqn:z12} -- \ref{eqn:z13}) immediately allow one to simultaneously 
solve for changes in both $\alpha$ and $(m_e/m_p)$. Of course, one further 
equation is needed to simultaneously constrain the evolution of all three constants.
Candidates include the HI 21cm and mm-wave molecular lines. The best of these 
are likely to be the HCO$^+$ lines since HCO$^+$ and OH column densities are 
found to show a strong correlation (extending over more than two orders of magnitude
in column density) both in the Galaxy \cite{liszt96} and out to $z \sim 1$; 
\cite{kanekar02}, this suggests that HCO$^+$ and OH are located in the same 
region of the molecular cloud. Since the HCO$^+$ line arises from a rotational 
transition, we have
\beq
\label{eqn:mm}
\frac{\Delta \nu_{HCO^+}}{\nu_{HCO^+}} = \frac{\Delta y}{y}
+ \frac{\Delta R_\infty}{R_\infty} \;\; ,
\eeq

Equations~(\ref{eqn:z12}), (\ref{eqn:z13}) and (\ref{eqn:mm}) 
all have the same dependence on the Rydberg constant $R_\infty$, which hence 
cancels out (note that $R_\infty$ itself depends on $\alpha$ 
through the relation $R_\infty \equiv m_e e^4/\hbar ^3 c$). One might thus 
combine HCO$^+$ absorption redshifts with those derived from the 18cm OH 
lines to provide the last equation needed to solve for the evolution of 
$\alpha$, $y$ and $g_p$. This would allow a simultaneous measurement 
of all three constants, which has been hitherto impossible. We note that 
it is also possible to use other OH $\Lambda$-doubled transitions to 
simultaneously constrain the variation of the above parameters; this 
is discussed elsewhere \cite{kanekar04}, for the ``main'' OH lines. It 
would also be interesting to carry out 
an analysis similar to that of Bekenstein \cite{bekenstein03} to test 
whether these analyses of the OH lines are also affected by the possibility 
that the Hamiltonians involved might vary with time (i.e. {\it dynamical} 
variability of the different parameters); this is beyond the scope of the 
present paper. It should also be pointed out that the present calculation is 
based on an analysis of the 
OH levels using perturbation theory \cite{dousmanis55}; more recent 
analyses \cite{brown79} use the ``effective Hamiltonian'' approach, giving
rise to higher order effects. While it would be interesting to attempt 
the present calculation in the latter framework, we emphasise that it 
should only result in small changes to the coefficients in 
equations~(\ref{eqn:temp1} -- \ref{eqn:temp3}) and does not affect the 
validity of our approach. Finally, while other ``Lambda-doubled'' systems 
could, in principle, be used for a similar analysis, none of these have 
multiple transitions detected in astrophysical objects (to the best of 
our knowledge). While it would be interesting to carry out searches for 
these other transitions, we suspect that OH is yet likely to prove the 
best candidate because of the strength of its multiple lines.

\section{Application to the $z = 0.6846$ absorber towards B0218+357}
\label{sec:0218}

While the above analysis shows that one can use the four 18cm OH lines 
(in conjunction with the HCO$^+$ transition) to constrain the evolution 
of three separate fundamental parameters, the weaker, satellite 1612~MHz 
and 1720~MHz lines have so far not been detected at cosmological distances. 
Observations are currently being scheduled to carry out deep searches for these 
lines in the four known OH absorbers at intermediate redshift 
\cite{chengalur99,kanekar02,kanekar03}. For the present, we will instead use the 
detected 1665~MHz and 1667~MHz transitions (i.e. equations~(\ref{eqn:temp1}) 
and (\ref{eqn:temp2})) along with the HI 21cm and millimetre (HCO$^+$) lines 
in the $z = 0.6846$ absorber towards B0218+357 to estimate changes in the 
fundamental parameters (assuming that the lines arise in the same gas cloud). 
Since the HI 21cm frequency arises from a hyperfine split and is hence 
proportional to $g_p y \alpha^2 R_\infty$, we have
\beq
\label{eqn:HI}
\frac{\Delta \nu_{21}}{\nu_{21}} = \frac{\Delta y}{y}
+ 2 \frac{\Delta \alpha}{\alpha} + \frac{\Delta g_p}{g_p} 
+ \frac{\Delta R_\infty}{R_\infty}
\eeq

\noi Equations~(\ref{eqn:temp1}), (\ref{eqn:temp2}), (\ref{eqn:mm}) and 
(\ref{eqn:HI}) can now be solved to measure changes in $\alpha$, $g_p$ and 
$m_e/m_p$. The HI 21cm redshift is $z_{HI} = 0.684676 \pm 0.000005$ 
\cite{carilli00}, while that of the HCO$^+$ absorption lines is 
$z_{HCO^+} = 0.684693 \pm 0.000001$ \cite{wiklind97}. Our new GMRT OH absorption 
spectra towards B0218+357 \cite{kanekar03} yield the following redshifts 
for the sum and difference of the 1665~MHz and 1667~MHz line frequencies 
: $z_{sum} = 0.684682 \pm 0.0000056$ and $z_{diff} = 0.685780 \pm 
0.0067$. A simultaneous solution of equations~(\ref{eqn:temp1}), 
(\ref{eqn:temp2}), (\ref{eqn:mm}) and (\ref{eqn:HI}) then yields 
$(\Delta \alpha / \alpha) = (-0.38 \pm 2.2) \times 10^{-3}$, 
$(\Delta y/ y) = (-0.27 \pm 1.6) \times 10^{-3}$ and $(\Delta g_p/ g_p) 
= (-0.77 \pm 4.2) \times 10^{-3}$. Since the error on $z_{diff}$ is far higher 
than the other errors, this dominates the errors on the above estimates and results in
relatively uninteresting upper limits on changes in the three constants. We emphasise,
however, that, to the best of our knowledge, this is the first time that a simultaneous
constraint on the variation of these three fundamental parameters has 
been obtained in a cosmologically distant object. Further, the 
weakness of the constraint arises from the relatively small difference between 
the frequencies of the main lines; this is clearly not a fundamental limitation 
but depends entirely on the sensitivity of the observations.

Next, if we {\it assume} that $y \equiv m_e/m_p$ is constant, 
equations~(\ref{eqn:temp1}) and (\ref{eqn:mm}) yield $\lsb \dal \rsb 
= - 5.7 \pm 3.0 \times 10^{-6}$. This is more than $3\sigma$ deviant from
the estimate $\lsb 2\Delta \alpha/\alpha \rsb = 1 \pm 0.3 \times 10^{-5}$ of
Carilli et al. (2000) \cite{carilli00}. Since the latter analysis assumed
that $g_p$ was constant, the difference between the two estimates implies
either that the assumptions that $y$ and/or $g_p$ are constant is
unjustified or that systematic velocity offsets do exist between the three
species.

Further progress can be made if one of the three parameters, $\alpha$, $g_p$ 
and $y$ is assumed to not change with time (while retaining the assumption that 
velocity offsets are not significant). We can then avoid having to use the 
equation governing the difference between the OH redshifts and can thus obtain a 
far stronger limit on changes in the remaining two quantities. For example, if 
we assume (as is often done, e.g. \cite{carilli00}) that $g_p$ remains constant, 
the HI 21cm and mm-wave molecular line redshifts imply $\lsb \dal  \rsb = (5 \pm 1.5)\times 10^{-6}$ 
\cite{carilli00}. Combining equations~(\ref{eqn:temp1}) and (\ref{eqn:mm}) 
then yields $\lsb \Delta y / y\rsb = (7.8\pm 2.4) \times 10^{-6} $.
Similarly, if we assume that $y \equiv m_e/m_p$ is constant, 
equations~(\ref{eqn:temp1}) and (\ref{eqn:mm}) yield $\lsb \dal \rsb = 
(- 5.7 \pm 3.0) \times 10^{-6}$ and $\lsb \Delta g_p /g_p\rsb = (2.2 \pm 
0.67) \times 10^{-6}$. Finally, if $\alpha$ remains unchanged, we obtain 
$\lsb \Delta y/y \rsb = (4.2 \pm 2.2) \times 10^{-6}$ and $\lsb \Delta g_p/g_p 
\rsb = (1 \pm 0.3) \times 10^{-5}$. The above limits on $\Delta y/y$ are a 
factor of $\sim 4$ stronger than the best earlier limits on changes in this 
quantity \cite{ivanchik03}. It is very interesting that all three cases 
result in a higher than $3\sigma$ significance for the variation 
of at least one of the parameters. This implies either that one (or more) 
of these parameters indeed varies with cosmological time or that systematic
motions between the three species cause the above uncertainties (which only 
include measurement errors) to be under-estimated.

Finally, we note that, while the OH 1667~MHz and 1665~MHz, HCO$^+$ and 
HI lines have been detected in four absorbers at intermediate redshifts, two 
of the absorbers (PKS1413+135 and B2~1504+377) are believed to have velocity 
offsets between the HI and HCO$^+$ redshifts \cite{carilli00,wiklind96a}. 
The last system, at $z \sim 0.889$ towards PKS1830$-$21, does not at 
present have OH data of sufficiently high quality to carry out the above 
analysis.

In summary, we have demonstrated a new technique to simultaneously measure 
the evolution of the three fundamental constants $\alpha$, $g_p$ and $m_e/m_p$, 
using 18cm OH absorption lines in conjunction with one additional transition, 
(which could be an HCO$^+$ mm-wave line). At present, only the 1665~MHz and 1667~MHz 
``main'' OH lines have been discovered at cosmological distances; we have 
used these line redshifts in conjunction with those of HI 21cm absorption and 
millimetre-wave molecular lines to constrain the variation of $y$, $g_p$ and 
$\alpha$ between $z = 0.6846$ and today.  We argue that one (or more) of 
the parameters $\alpha$, $y$ and $g_p$ must vary with cosmological time, 
unless systematic velocity offsets exist between the above three species. 
The constraints placed on changes in the parameter $y \equiv m_e/m_p$ 
(assuming that either $\alpha$ or $g_p$ are constant) are a factor of 
$\sim 4$ stronger than earlier limits on variations in this parameter.

\begin{acknowledgments}
	We are grateful to Rajaram Nityananda for very useful discussions
on the energy levels of the OH ground state.
\end{acknowledgments}

\bibliography{ms}

\end{document}